\documentclass[reprint,superscriptaddress,aps,prx,longbibliography,nofootinbib]{revtex4-1}

\usepackage{amsmath}
\usepackage{amssymb}
\usepackage{amsthm}
\usepackage{mathtools}
\usepackage{mathdots}
\usepackage{amsfonts}
\usepackage{bbm}
\usepackage{xr}
\usepackage{bbold}
\usepackage{newpxtext}
\usepackage{epsfig,color,dsfont,upgreek,physics}
\usepackage{mathrsfs}
\usepackage{multirow}
\usepackage{graphicx}
\usepackage{dcolumn} 
\usepackage{bm} 
\usepackage{float} 
\usepackage[driverfallback=dvipdfm]{hyperref} 
\usepackage{longtable}
\usepackage{ulem}  
\usepackage{enumerate}
\usepackage{comment}
\allowdisplaybreaks

\renewcommand\[{\begin{equation}}
\renewcommand\]{\end{equation}}

\def\thefootnote{*}\footnotetext{J. H. and D. S. contributed equally to this work.}

\begin{document}

\title{Superconducting diode efficiency from singlet-triplet mixing in disordered systems}

\author{Jaglul Hasan\thefootnote{}}
\affiliation
{Department of Physics, University of Wisconsin-Madison, Madison, Wisconsin 53706, USA}

\author{Daniel Shaffer\thefootnote{}}
\affiliation
{Department of Physics, University of Wisconsin-Madison, Madison, Wisconsin 53706, USA}

\author{Maxim Khodas}
\affiliation{The Racah Institute of Physics, The Hebrew University of Jerusalem, Jerusalem 91904, Israel}

\author{Alex Levchenko}
\affiliation
{Department of Physics, University of Wisconsin-Madison, Madison, Wisconsin 53706, USA}

\begin{abstract}
The superconducting diode effect (SDE) -- the nonreciprocity of the critical current in a bulk superconductor -- has garnered significant attention due to its potential applications in superconducting electronics. However, the role of disorder scattering in SDE has rarely been considered, despite its potential qualitative impact, as we demonstrate in this work. We investigate SDE in a disordered Rashba superconductor under an in-plane magnetic field, employing a self-consistent Born approximation to derive the corresponding Ginzburg-Landau theory. Our analysis reveals two surprising effects. First, in the weak Rashba spin-orbit coupling (SOC) regime, disorder can reverse the direction of the diode effect, indicated by a sign change in the superconducting diode efficiency coefficient. Second, in the strong Rashba SOC regime, disorder becomes the driving mechanism of SDE, which vanishes in its absence. In this case, we show that disorder-induced mixing of singlet and triplet superconducting orders underlies the effect.
\end{abstract}

\date{April 3, 2025}

\maketitle


\section{Introduction}

In the absence of both time reversal and inversion symmetries, transport in bulk superconductors becomes nonreciprocal \cite{NadeemFuhrerWang23}: critical currents \(J_{c+}\) and \(J_{c-}\) running in opposite directions are not precisely opposite $|J_{c+}|\neq|J_{c-}|$ -- the so-called intrinsic superconducting diode effect (SDE)  that has been observed in several recent experiments \cite{AndoYanaseOno20, LinScheurerLi22, BauriedlParadiso22, HouMoodera23,  KealhoferBalentsStemmer23, GengBergeretHeikkila23, YunKim23, LeLin24, AsabaYanaseMatsuda24, JamesLeo2024NatMat, LiuIwasa24, WanDuan24, Ingla-AynesMoodera24}. Related effects have also been seen in superconducting heterostructures like Josephson junctions, including the Josephson diode (JDE) \cite{WakatsukiNagaosa17, LotfizadehShabani24, Moll24, YerinGiazotto24} and anomalous Josephson (AJE) effects \cite{SzombatiNadjPergeKouwenhoven16, BaumgartnerManfraStrunk22, ReinhardtManfraGlazmanStrunk24}; as well as extrinsic SDE due to boundary effects \cite{SwartzHart67}, which includes the vortex diode effect \cite{CerbuVondel13, GutfreundAnahory23, HouMoodera23}.

Consequently, and building on some early theoretical work \cite{LevitovNazarovEliashberg85, GeshkenbelhLarkin86,  edelstein_ginzburg_1996, Sigrist98, Agterberg2005,VodolazovPeeters05,  Buzdin05, HuWuDai07, ReynosoAvignon08}, a lot of effort has been directed to develop the theory of nonreciprocal superconductivity in general \cite{Agterberg2011,DavydovaFu22, DavydovaFu24,  ZhangJiang22, LiuAndreevSpivak24, KochanZutic23, MonroeZutic24, PekertenZutic24, NunchotYanase24_2}, and of SDE in particular \cite{DaidoYanase22, YuanFu22, HeTanaka2022, IlicBergeret22, KapustinRadzihovsky22, ZinklHamamotoSigrist22, HasanLevchenko23, HasanShafferKhodasLevchenko24, OsinLevchenkoKhodas24, ShafferChichinadzeLevchenko24, GaggioliGeshkenbein24, NunchotYanase24, DaidoYanase23, KotetesAndersen23, BanerjeeScheurer24, BanerjeeScheurer24alt, ZhangNeupert24, SimKnolle24, ChakrabortyBlackSchaffer24, HuLaw24}. However, not much attention has been payed to disorder effects on the SDE, despite their ubiquity in real systems. Only a few works have studied the role of disorder on SDE numerically, in the context of helical superconductivity (SC) in a 2D superconductor with Rashba spin-orbit coupling (SOC) in an in-plane magnetic field \cite{IlicBergeret22, IkedaDaidoYanase22, IlicBergeret24}. In particular, these studies have been limited to the case of large SOC in which interband pairing between the two SOC-split bands can be neglected, and considered only the weak disorder \cite{IlicBergeret22, IkedaDaidoYanase22} or strong disorder (diffusive) \cite{IlicBergeret24} limits. Moreover, the self-consistency condition imposed on the order parameter assumed in \cite{IlicBergeret22} is only valid in the limit of very strong SOC when corrections due to interband pairing, as well as to the form of the intraband terms, can be neglected, as discussed in \cite{HasanShafferKhodasLevchenko24}.

In this work, we revisit the problem of SDE in disordered helical Rashba SCs with \(s\)-wave singlet pairing interactions using a self-consistent Born approximation. In contrast to earlier works, we consider an arbitrary SOC strength and treat the SC order parameter self-consistently, properly including interband pairing. Limiting our attention to small magnetic fields and temperatures close to the critical temperature allows us to perturbatively derive the Ginzburg-Landau (GL) theory. We then obtain numerical and analytical expressions for the superconducting diode efficiency \(\eta=(J_{c+}+J_{c-})/(J_{c+}-J_{c-})\).

We find two surprising effects, in the limit of weak and strong SOC, respectively. First, in the weak SOC limit we find that sufficiently strong disorder can reverse the sign of \(\eta\). Second, in the strong SOC limit, in which \(\eta\) was shown to vanish in \cite{IlicBergeret22} due to an accidental approximate symmetry \cite{HasanShafferKhodasLevchenko24}, we find that disorder can \emph{induce} a finite \(\eta\). 
We attribute this effect to a disorder-induced triplet pairing component, which is well-known to occur in disordered noncentrosymmetric SCs \cite{BauerSigrist12, Samokhin08, Edelstein05, *Edelstein08}. This effect leads to a nonmonotonic dependence of \(\eta\) on disorder and SOC strengths. These findings illustrate the general importance of considering disorder effects on SDE.

The paper is organized as follows. We first outline the details of our model and the diagrammatic technique in Sec. \ref{SecII}. Next Sec. \ref{SecIII} contains the derivation of the GL theory with disorder. We then present the results of our calculations in Sec. \ref{SecIV}, including analytical and numerical values of \(\eta\) and \(J_{c\pm}\) in various limits, focusing in particular on the weak and strong SOC limits. Finally, we discuss the implications of our results in Sec. \ref{SecV}.

\section{Model and Technique}\label{SecII}

\subsection{Single-particle model}\label{SecIIA}

The single-particle Hamiltonian of the two-dimensional (2D) clean noncentrosymmetric metal with Rashba SOC has the form \cite{Rashba1984}
\begin{equation} \label{eq:III1}
H^{(0)}(\mathbf{p})=\frac{\mathbf{p}^2}{2 m}+\alpha_R(\mathbf{p} \times \mathbf{c}) \cdot \bm{\sigma}\,.
\end{equation}
Here $\mathbf{p}$ is the momentum, \(m\) is the electron effective mass, $\mathbf{c}$ is a unit vector perpendicular to the 2D metal, $\mathbf{\sigma}$ is the Pauli spin matrix-vector, and $\alpha_{R}$ is the strength of the Rashba spin-orbit (SO) interaction with dimensions of velocity (here and below we set $\hbar=1$ and $k_B=1$).
The SOC lifts the spin degeneracy of the conduction electrons, giving rise to two energy bands of positive and negative helicities \(\lambda=\pm1\)
with energies $\epsilon_{\lambda}(p)=$ $\frac{p^2}{2 m} +\lambda \alpha_{R} p$. 

We consider the effect of an external uniform magnetic field \(\mathbf{B}\) via the Zeeman coupling term
\[V^{\text{(Z)}}=\mu_B \bm{\sigma}\cdot \mathbf{B}\]
where $\mu_B$ is the Bohr magneton. We introduce the field strength \(\mathbf{h}=\mu_B\mathbf{B}\) and treat \(h=|\mathbf{h}|\) as a perturbation.

We include disorder as a short-ranged potential of impurities placed at randomly distributed points $\mathbf{R}_i$ with concentration $n_{\text{imp}}$: 
\begin{equation}\label{eq:imp_pot}
V^{(\text{imp})}(\mathbf{r})=\sum_i U \delta\left(\mathbf{r}-\mathbf{R}_i\right).
\end{equation}
The elastic scattering time $\tau$ is given by $\tau^{-1}=2 \pi \nu n_{\text{imp}}|U|^2$, where $\nu=m / 2 \pi$ is the density of states on the 2D Fermi level.

Though an electron may scatter on a single impurity and cause a transition between the bands, the one-particle Green's function averaged over impurity positions is diagonal in the helicity index. Nevertheless, all possible scattering channels contribute comparably into the impurity ladder, which means that intraband and interband transitions contribute on equal footing.

Working at finite temperature \(T\), there are four independent dimensionless parameters in the model:
\begin{align}
\delta=\frac{\alpha_R }{v_F}, \quad \kappa=\frac{\alpha_R p_F}{T}, \quad 
\gamma=\frac{h}{T}, \quad \varkappa=2\alpha_R p_F\tau\,,
\end{align}
where \(p_F=\sqrt{2m\mu}\) is the Fermi momentum at chemical potential \(\mu\), and \(v_F=p_F/m\) is the Fermi velocity (in the absence of SOC). We assume that the parameters $\delta$ and \(\gamma\) are very small and work to first order in both, but the parameter $\kappa$ and $\varkappa$ are arbitrary. 

\subsection{Diagrammatic technique}\label{SecIIB}

The difference between the diagrammatic technique used here and the conventional one is that due to the SOC the key ingredients (vertices, propagators, and, more importantly, the impurity ladders) acquire a specific spin structure similar to that in quantum electrodynamics \cite{edelstein_ginzburg-landau_2021}.
From the Hamiltonian in Eq. (\ref{eq:III1}), it follows that the thermal Green function of noninteracting electrons in the absence of an external field and impurity potential is given by 
\begin{equation}
G^{(0)}_{ab}\left(i \omega_n, \mathbf{p}\right)  =\sum_{\nu= \pm} \Pi_{ab}^{(\nu)}(\mathbf{p}) G_{(\nu)}\left(i \omega_n, p\right),
\end{equation}
with \(a,b\) being spin indices. Here
\begin{equation}
\Pi^{( \pm)}_{ab}(\mathbf{p})  =\frac{1}{2}\left(\delta_{ab} \pm \frac{(\mathbf{p} \times \mathbf{c}) \cdot \bm{\sigma}_{ab}}{|\mathbf{p} \times \mathbf{c}|}\right),
\end{equation}
is the projection operator onto a state with a definite helicity with 
\begin{equation}\label{eq:GF_branch}
G_{(\nu)}\left(i \omega_n, p\right)=\left[i \omega_n-\xi_{(\nu)}(p)\right]^{-1}.
\end{equation}
and $\xi_{( \pm)}(p)=\epsilon_{ \pm}(p)-\mu$. This Green function is the fundamental tool for the diagram technique. 

In addition, it is convenient to introduce the reversed Green's function, $G_{ab}^{(\text{r})}\left(i \omega_n, \mathbf{p}\right)$ via the equation
\begin{equation}
-G_{ab}^{\text{t(0)}}\left(-i \omega_n,-\mathbf{p}\right)=g_{ac}^{\mathrm{t}} G_{cd}^{(0)(\text {r})}\left(i \omega_n, \mathbf{p}\right) g_{db}, 
\end{equation}
where where $g=i \sigma_y$ and the superscript $t$ denotes matrix transposition. Then
\begin{subequations}
\begin{align}
G_{ab}^{(0)(\mathrm{r})}\left(i \omega_n, \mathbf{p}\right) & =\sum_{\nu= \pm} \Pi_{ab}^{(\nu)}(\mathbf{p}) G_{(\nu)}^{(\mathrm{r})}\left(i \omega_n, p\right), \\
G_{(\nu)}^{(\mathrm{r})}\left(i \omega_n, p\right) & =\left[i \omega_n+\xi_{(\nu)}(p)\right]^{-1}.
\end{align}
\end{subequations}
In this representation, we have to keep track of the spinor structure of the Green's function and the explicit form of the velocity operator:
\begin{equation}
\mathbf{v}_{ab}(\mathbf{p})=i\left[H^{(0)}_{ab}(\mathbf{p}), \mathbf{r}\right]=\frac{\mathbf{p}}{m} \delta_{ab}+\alpha_{R}(\mathbf{c} \times \bm{\sigma})_{ab},
\end{equation}
which has a spin component alongside the usual band velocity part.

To include superconductivity, we assume s-wave pairing interactions:
\begin{equation}
H^{(\text{SC})}=\frac{\lambda_s}{4} \int \mathrm{d}^3 r\left[\psi_a^{\dagger}(\mathbf{r}) g_{ab} \psi_b^{\dagger}(\mathbf{r})\right]\left[\psi_c(\mathbf{r}) g_{cd}^{t} \psi_d(\mathbf{r})\right],
\end{equation}
with $\lambda_s$ being the pairing constant and $\psi_b(\mathbf{r})$ representing the electron field operator. The Gor'kov equations for the matrix Green's function
\begin{equation}
\hat{G}_{ab}=\left(\begin{array}{cc}
G_{ab}(i \omega_n, \mathbf{p}, \mathbf{q}) & F_{ab}(i \omega_n, \mathbf{p}, \mathbf{q}) \\
F_{ab}^{\dagger}(-i \omega_n, \mathbf{p}, \mathbf{q}) & -G_{ab}^{\mathrm{t}}(-i \omega_n, -\mathbf{p},-\mathbf{q})
\end{array}\right)
\end{equation}
have the standard form
\begin{equation}\label{eq:Gorkov-b}
\int \frac{\mathrm{d}^2 q'}{(2 \pi)^2} \hat{K}_{ac}\left(\mathbf{p}, \mathbf{q}'\right) \hat{G}_{cb}\left(\mathbf{q}', \mathbf{q}\right)=(2 \pi)^2 \delta(\mathbf{p}-\mathbf{q}) \delta_{ab} \hat{1},
\end{equation}
where the kernel is defined by 
\begin{widetext}
\begin{equation}
\hat{K}_{ab}(\mathbf{p}, \mathbf{q})=(2 \pi)^2 \delta(\mathbf{p}-\mathbf{q})\left(\begin{array}{cc}
i \omega_n \delta_{ab}-H_{ab}^{(0)}(\mathbf{p})-V^{\text{(Z)}}_{ab}-V^{\text{(imp)}}_{ab}(\mathbf{q}) & -\Delta_{ab}(\mathbf{p}-\mathbf{q}) \\
-\Delta_{ab}^{\dagger}(\mathbf{p}-\mathbf{q}) & i \omega_n \delta_{ab}+H_{ab}^{\mathrm{t}(0)}(-\mathbf{p})+V^{\text{t(Z)}}_{ab}+V^{\text{t(imp)}}_{ab}(\mathbf{q})
\end{array}\right).
\end{equation}
\end{widetext}
The self-consistent order parameter is defined through the trace of the anomalous part of the Green's function  
\begin{equation}\label{eq:Gorkov-d}
\Delta_{ab}(\mathbf{p})=-\frac{\lambda_s}{2} g_{ab} T \sum_{\omega_n} \int \frac{\mathrm{d}^2 q}{(2 \pi)^2} \operatorname{Tr}\left[F(i\omega_n, \mathbf{p}, \mathbf{q}) \cdot g^{\mathrm{t}}\right]
\end{equation}
Here $\omega_n=2\pi T(n+1/2)$ with $n\in\mathbb{Z}$ are fermion Matsubara frequencies, $V^{\text{t(imp)}}(\mathbf{q})$ is the impurity potential from Eq. (\ref{eq:imp_pot}) in momentum space.


\begin{figure}[t!]
\includegraphics[width=0.48\textwidth]{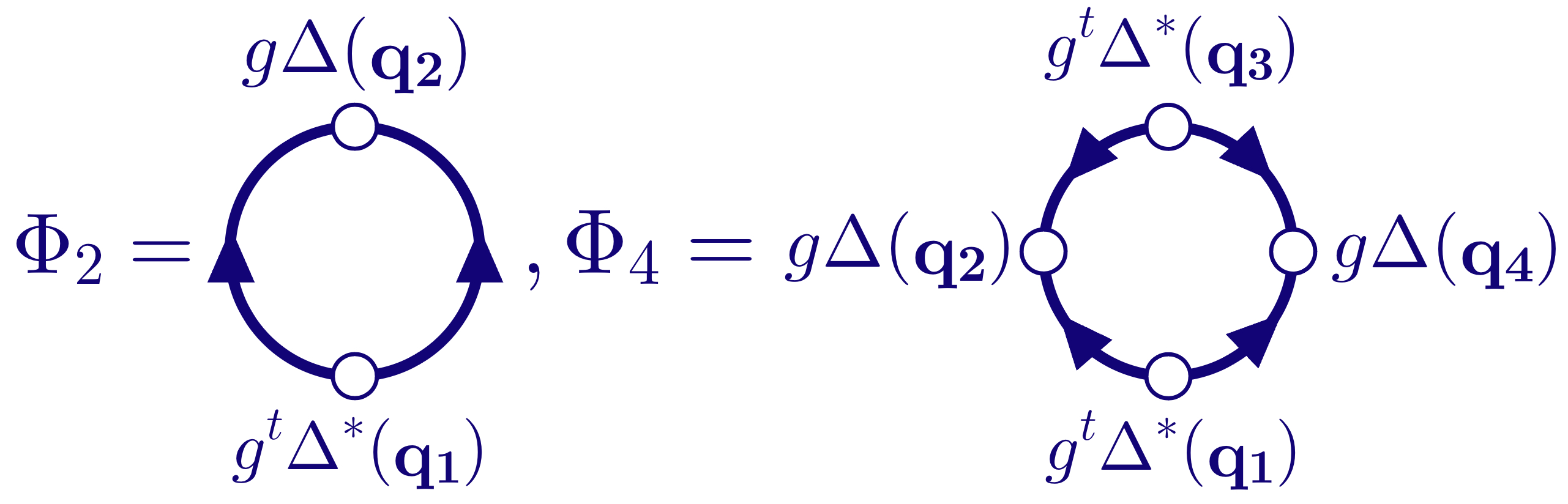} 
\caption{The mean-field loop diagrams for the GL functional Eq. \eqref{eq:IIIA8}. Here, thick clockwise fermion line of the $\Phi_2$ diagram denotes $G\left(i \omega_n, \mathbf{q_1},\mathbf{q_2}\right)$ while thick anticlockwise line denotes $-G^{\text{t}}\left(-i \omega_n, -\mathbf{q_1},-\mathbf{q_2}\right)$.}  \label{fig2}
\end{figure}

\section{Microscopic Calculation of GL Coefficients}\label{SecIII}

We obtain the GL equation to fourth order in \(\Delta(\mathbf{q})\) by iterating Eq. (\ref{eq:Gorkov-b}) to third order in $\Delta(\mathbf{q})$, and then substituting the result into Eq. (\ref{eq:Gorkov-d}) for self-consistency. 
This is equivalent to obtaining the stationary point 
\begin{equation}
\frac{\delta \Omega}{\delta {\Delta}^{*}(\mathbf{q})}=0,
\end{equation}
for the thermodynamic potential
\begin{equation}\label{eq:IIIA8}
\begin{aligned}
\Omega&=\frac{1}{|\lambda_s|} \int_{\mathbf{q}}|\Delta(\mathbf{q})|^2+\frac{1}{2}\Phi_2+\frac{1}{4} \Phi_4 \\&= \int_{\mathbf{q}}\left[\alpha(q) |\Delta\left(\mathbf{q}\right)|^2+\beta(q) |\Delta\left(\mathbf{q}\right)|^4\right],
\end{aligned}
\end{equation}
with the notation $\int_{\mathbf{q}}=\int \frac{\mathrm{d}^2 q}{(2 \pi)^2}$. For the purposes of obtaining the diode efficiency $\eta$, the GL expansion coefficients are further written in a series over the collective Cooper pair momentum  
\begin{equation}\label{eq:alpha-beta-q}
\alpha(q)=\sum_{n \geq 0} \alpha_n q^n, \quad \beta(q)=\sum_{n \geq 0} \beta_n q^n.
\end{equation}
The efficiency can be expressed via $\alpha_n$ and $\beta_n$ [see Eq. \eqref{etaGL}]. It is crucial to note that odd powers of $\alpha_n$ and $\beta_n$ are only possible in the presence of the field. 

\begin{figure}
\includegraphics[width=0.35\textwidth]{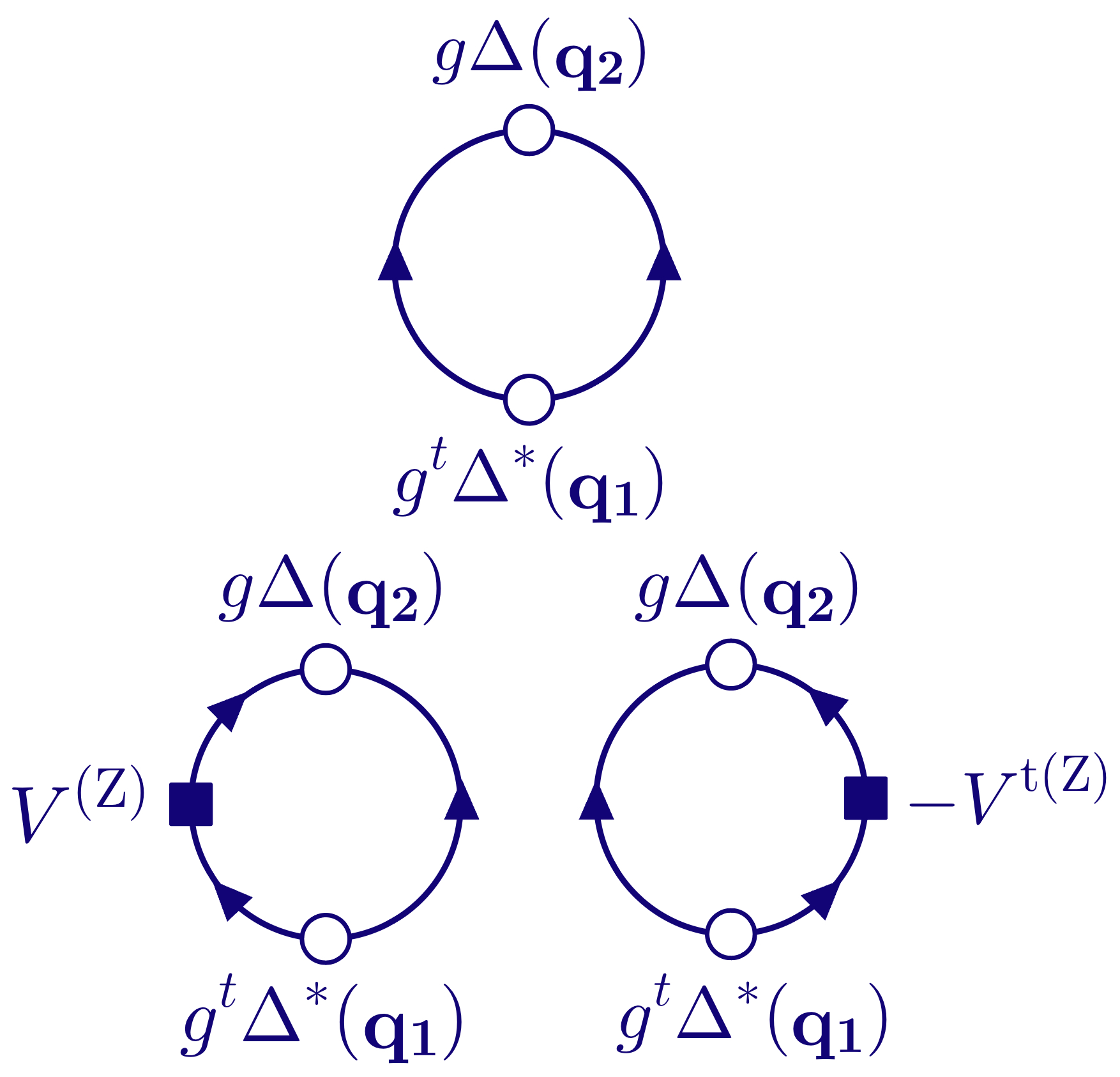} 
\caption{Diagrammatic representation for $\Phi_{\mathrm{2}}$ expanded in a series in external magnetic field. The solid square denotes $V^{\text{(Z)}}$ if it is placed on a clockwise solid line, whereas being placed on an anticlockwise line it denotes $-V^{\text{t(Z)}}$.}  \label{fig3}
\end{figure}

\begin{figure}
\includegraphics[width=0.48\textwidth]{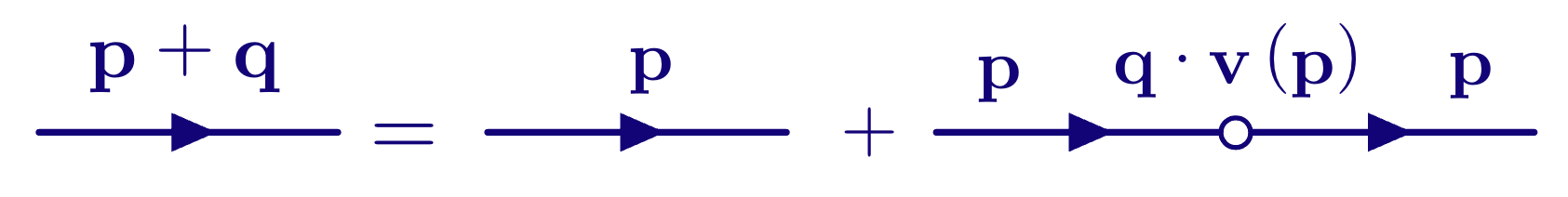} 
\caption{Diagram expansion of the Green's function $G^{(0)}_{\alpha\beta}(i \mathbf{\omega_n},\mathbf{p}+\mathbf{q})$ in the collective momentum $\mathbf{q}$.}  \label{fig4}
\end{figure}

In the expression for $\Omega$, $\Phi_2$ and $\Phi_4$ are the quadratic and quartic functionals of $\Delta(\mathbf{q})$, respectively, defined diagrammatically in Fig. \ref{fig2}. To make further analytical progress, we expand the diagrams in Fig. \ref{fig2} into a series in the small external field $\mathbf{B}$. For example, the functional $\Phi_2$ transforms into the sum of diagrams represented by Fig. \ref{fig3}. Unlike Fig. \ref{fig2}, thin solid lines in Fig. \ref{fig3} correspond to the normal system subject to the field of impurity potential (prior to disorder averaging) but without the interaction with the magnetic field. 
The propagator $G^{(0)}_{ab}(i \mathbf{\omega}_n,\mathbf{p}+\mathbf{q})$ is further expanded in a power series in the Cooper pair momentum $\mathbf{q}$ as in Fig. \ref{fig4}.

\begin{figure}
\includegraphics[width=0.35\textwidth]{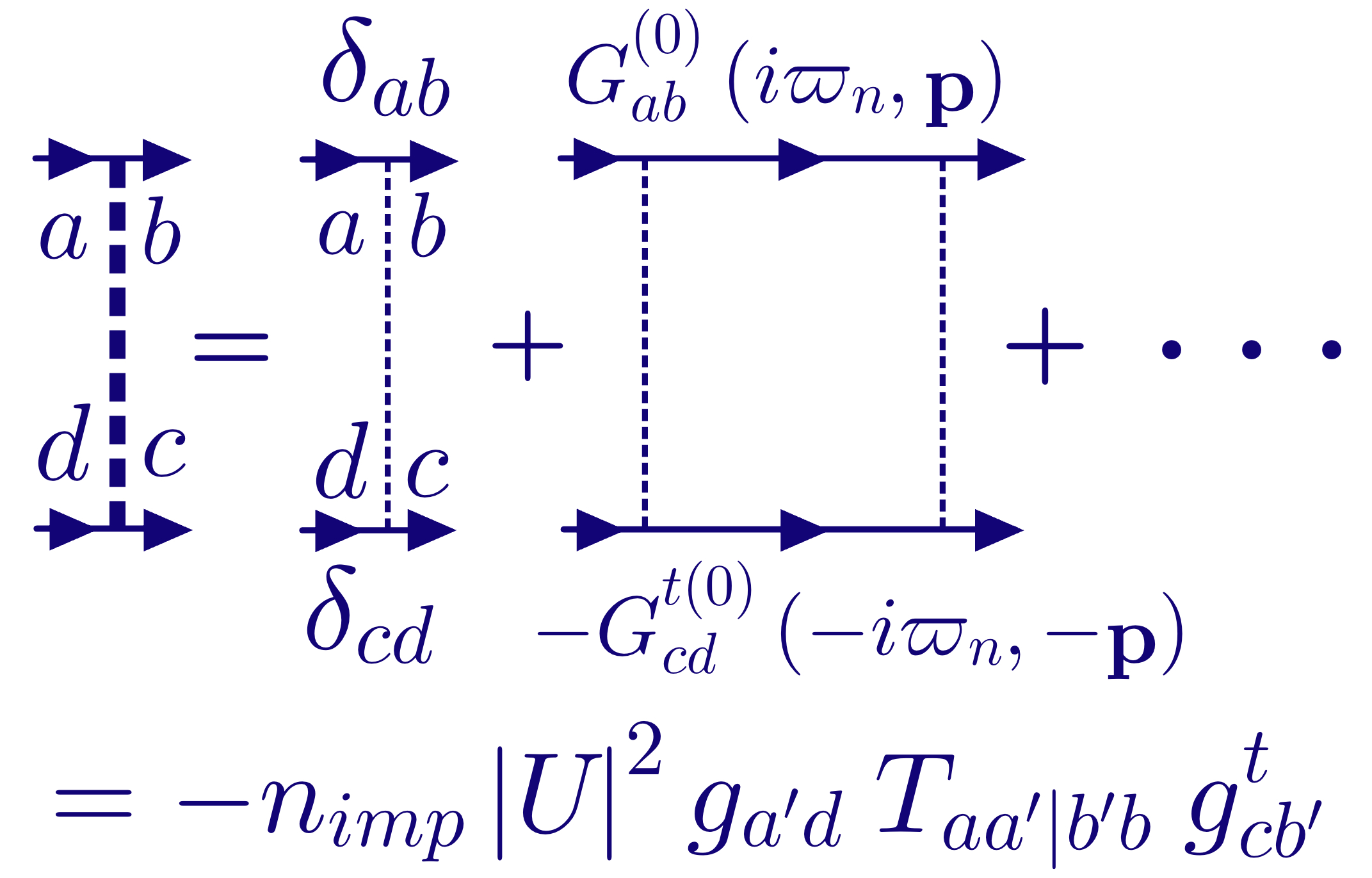} 
\caption{The impurity ladder represented by a thick dashed line.}  \label{fig5}
\end{figure}

\begin{figure}
\includegraphics[width=0.35\textwidth]{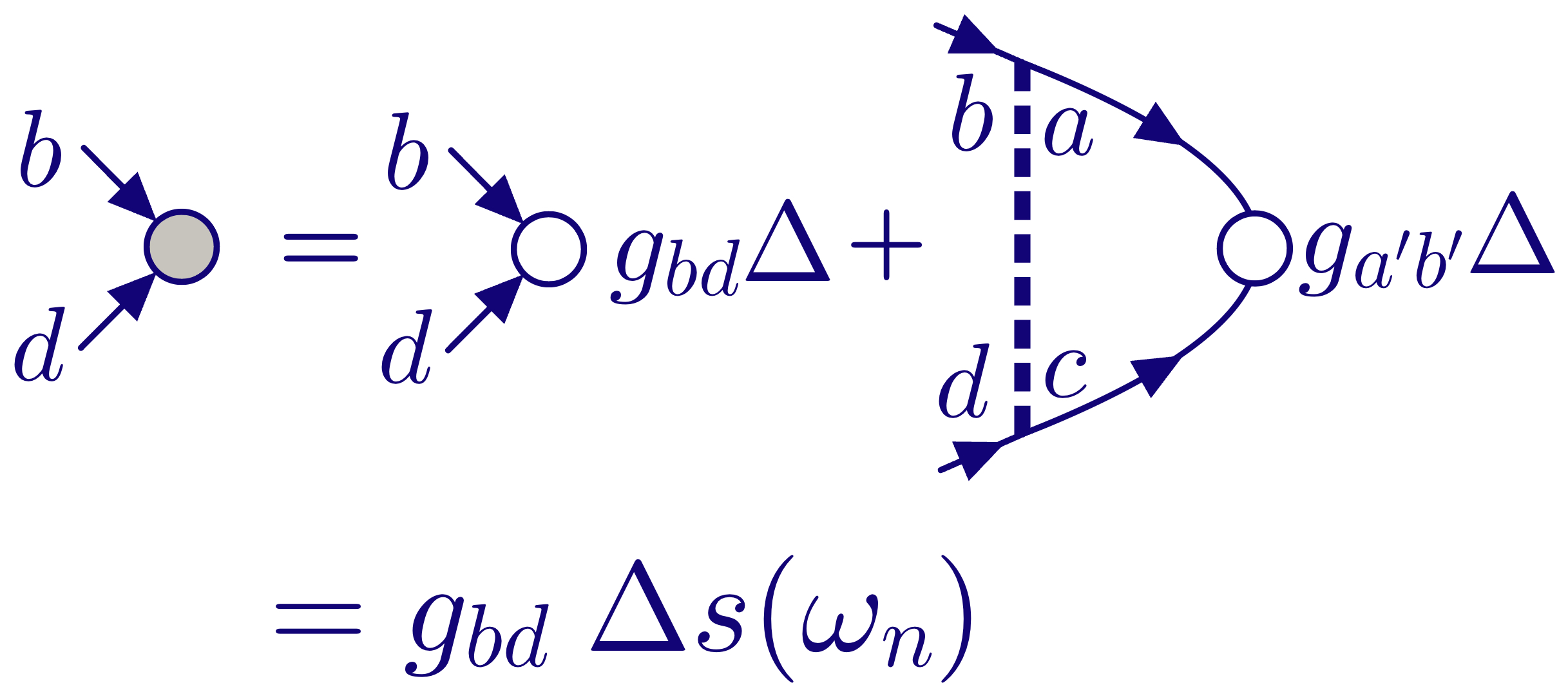} 
\caption{The equation for the impurity-renormalized order parameter represented by a solid gray circle. The argument of the order parameter is dropped here.}  \label{fig6}
\end{figure}

After averaging over disorder realizations, translation invariance is restored at the expense of introducing impurity lines, shown as thin dashed lines, to the diagrammatics, contributing $n_{\text{imp}}|U|^2$ when they connect two clockwise or two anticlockwise fermion lines; when an impurity line connects a clockwise fermion line with an anticlockwise fermion line, it contributes $-n_{\text{imp}}|U|^2$. The sum of impurity insertions (without intersections) onto a specific solid line results in the impurity renormalization of the frequency of the associated Green's function:
\begin{equation}
i\omega_n \rightarrow i \varpi_n=i \omega_n s\left(\omega_n\right), \quad s\left(\omega_n\right)=1+\left(2 \tau\left|\omega_n\right|\right)^{-1}.
\end{equation}

In disordered superconductors, the diagrams additionally contain impurity ladders represented by thick dashed lines defined diagrammatically in Fig. \ref{fig5} and impurity-renormalized order parameter represented by a solid gray circle shown in Fig. \ref{fig6}. The matrix $\widehat{T}\left(\omega_n\right)$ in Fig. \ref{fig5} is given by
\begin{subequations}
\begin{align}\label{eq:Tmatrix}
T_{ad \mid cb}\left(\omega_n\right)=&\frac{1}{2}\left[\frac{\delta_{ad} \delta_{cb}}{1-w\left(\omega_n\right)}+\frac{(\mathbf{c} \cdot \bm{\sigma})_{ad}(\bm{c} \cdot \mathbf{\sigma})_{cb}}{1-u\left(\omega_n\right)}\right.\nonumber\\&\left.+\frac{(\mathbf{c} \times \bm{\sigma})_{ad}^i(\mathbf{c} \times \bm{\sigma})_{cb}^i}{1-v\left(\omega_n\right)}\right],
\end{align}
where $w\left(\omega_n\right)$, $u\left(\omega_n\right)$ and $v\left(\omega_n\right)$ are given by
\begin{equation}
w\left(\omega_n\right)=\frac{1}{2 \tau\left|\varpi_n\right|}=\frac{1}{2 \tau\left|\omega_n\right| s\left(\omega_n\right)}=\frac{1}{2 \tau\left|\omega_n\right|+1},
\end{equation}
\begin{equation}
u\left(\omega_n\right)=\frac{2 \tau\left|\varpi_n\right|}{\left[2 \tau\left|\varpi_n\right|\right]^2+\left(2 \alpha_R p_F \tau\right)^2}=\frac{2 \tau\left|\omega_n\right|+1}{\left(2 \tau\left|\omega_n\right|+1\right)^2+\varkappa^2},
\end{equation}
\begin{equation}
v\left(w_n\right)=\frac{1}{2}\left[w\left(\omega_n\right)+u\left(\omega_n\right)\right].
\end{equation}
\end{subequations}

Now that we have established the full diagrammatic technique, the skeleton diagrams of a clean superconductor can be transformed to include impurity ladders, impurity lines, and impurity-dressed gap functions due to impurity scattering. We have cataloged all diagrams contributing to $\alpha_n$ and $\beta_n$ in the Ginzburg-Landau functional up to fourth order in the Cooper pair momentum $\mathbf{q}$. These diagrams are presented order by order in the supplemental file \cite{SM}.

To gain insight into the essential physics behind all diagrammatic contributions, we explicitly evaluate two diagrams: one for a clean superconductor, represented by the skeletal diagram in Fig. \ref{fig12}$(a)$, and one for a disordered superconductor, which includes an impurity ladder, shown in Fig. \ref{fig12}$(b)$. These evaluations are presented in Appendices \ref{AppA} and \ref{AppB}, respectively.

In the strong SOC limit $\kappa\gg1$ \cite{HasanShafferKhodasLevchenko24}, the diagram in Fig. \ref{fig12}$(a)$ remains diagonal in the helical indices. In contrast, the impurity-ladder diagram in Fig. \ref{fig12}$(b)$ is not diagonal in the helical basis, leading to mixing between the two helical bands and inducing interband pairing. Consequently, this mixing results in singlet-triplet coupling of the superconducting order parameter.


\section{Results}\label{SecIV}

The analytical computations of all diagrams, tailored to the limit near the critical temperature $T_c$, yield the following expressions for the coefficients of $\alpha(q)$ in Eq. \eqref{eq:alpha-beta-q}:
\begin{widetext}
\begin{subequations}
\begin{align}
& \alpha_0=-\nu\left(\frac{T_c-T}{T_c}\right),\\
& \alpha_1=-2 \nu \alpha_R  (\left[\mathbf{h} \times \hat{\mathbf{q}}\right] \cdot \mathbf{c}) \sum_{\omega_n>0} \frac{2\pi T}{\omega_n^2} \cdot \frac{\tau \varkappa^2}{4 \tau \omega_n\left(2 \tau \omega_n+1\right)^2+\varkappa^2\left(4 \tau \omega_n+1\right)},\\
& \alpha_2=\frac{1}{4} \nu v_F^2 \; 2\pi T \sum_{\omega_n>0} \frac{\tau}{\omega_n^2\left(2 \tau \omega_n+1\right)},\\
& \alpha_3=\frac{1}{4} \nu \alpha_R v_F^2 (\left[\mathbf{h} \times \hat{\mathbf{q}}\right] \cdot \mathbf{c}) 2\pi T \sum_{\omega_n>0} \frac{ \tau^2 \varkappa^2\left[f_1(\tau \omega_n)+ \varkappa^2 f_2(\tau \omega_n)+\varkappa^4 f_3(\tau \omega_n)\right]}{\omega_n^3(2 \tau \omega_n+1)^3\left[(2 \tau \omega_n+1)^2+\varkappa^2\right]\left[4 \tau \omega_n(2 \tau \omega_n+1)^2+\varkappa^2(4 \tau \omega_n+1)\right]^2},\\
& \alpha_4=-\frac{1}{16} \nu v_F^4 \; 2\pi T \sum_{\omega_n>0} \frac{\tau^2\left(3\omega_n\tau+1\right)}{\omega_n^3\left(2 \tau \omega_n+1\right)^3},
\end{align}
\end{subequations}
where $\hat{\mathbf{q}}$ is the unit vector along the direction of $\mathbf{q}$. The dimensionless functions are defined as follows: 
$f_1(z)=8 z(1+2 z)^4\left(3+20 z+30 z^2\right)$, $f_2(z)= 4(1+2 z)^2\left(1+18 z+68 z^2+68 z^3\right)$ and $f_3(z)=4+45 z+128 z^2+96 z^3$.
The coefficients of $\beta(q)$ up to the second order in $q$ are found in the form:
\begin{subequations}
\begin{align}
& \beta_0=\frac{1}{4} \nu \sum_{\omega_n>0} \frac{2\pi T}{ \omega_n^3}=\frac{7\zeta(3) \nu}{16 \left(\pi T\right)^2},\\
& \beta_1= \nu \alpha_R (\left[\mathbf{h} \times \hat{\mathbf{q}}\right] \cdot \mathbf{c}) 2\pi T \sum_{\omega_n>0} \frac{ \tau\varkappa^2\left[d_1(\tau \omega_n)+ \varkappa^2 d_2(\tau \omega_n)+\varkappa^4 d_3(\tau \omega_n)+\varkappa^6 d_4(\tau \omega_n)\right]}{\omega_n^4\left[\varkappa^2+(1+2 \tau \omega_n)^2\right]^2\left[4 \tau \omega_n(1+2 \tau \omega_n)^2+\varkappa^2(1+4 \tau \omega_n)\right]^2},\\
& \beta_2= -\frac{1}{4} \nu v_F^2 \; 2\pi T \sum_{\omega_n>0} \frac{\tau\left(3\omega_n\tau+1\right)}{\omega_n^4\left(2 \tau \omega_n+1\right)^2},
\end{align}
\end{subequations}
where $d_1(z)=16 z(1+2 z)^4\left(1+4 z+5 z^2\right)$, $d_2(z)= 2(1+2z)^2\left(1+24 z+92 z^2+104 z^3\right)$, $d_3(z)=4+53z+176z^2+176z^3$ and $d_4(z)=2\left(1+6z\right)$.
\end{widetext}

We can now compute the Matsubara sums above (see Appendix \ref{AppC} for details) and use the following formula, derived previously in Ref. \cite{HasanShafferKhodasLevchenko24}, for the calculation of the critical currents and the supercurrent diode efficiency:
\begin{subequations}
\begin{equation}
\begin{aligned}
J_{c\pm}=&\frac{2\left(2 \alpha_2 \alpha_3 \beta_0-4 \alpha_1 \alpha_4 \beta_0-\alpha_2^2 \beta_1+\alpha_1 \alpha_2 \beta_2\right) \alpha_0^2}{9 \alpha_2^2 \beta_0^2}\\
&\pm\frac{4 \sqrt{\alpha_2} (-\alpha_0)^{\frac{3}{2}}}{3 \sqrt{3} \beta_0},
\end{aligned}
\end{equation}
\begin{equation}
\eta=\frac{2 \alpha_2 \alpha_3 \beta_0-4 \alpha_1 \alpha_4 \beta_0-\alpha_2^2 \beta_1+\alpha_1 \alpha_2 \beta_2}{2\sqrt{3} \alpha_2^{\frac{5}{2}}\beta_0}\sqrt{-\alpha_0}.\label{etaGL}
\end{equation}
\end{subequations}
We also assume for simplicity that the in-plane magnetic field $\mathbf{h}$ is perpendicular to the direction of the superflow.

\subsection{Coefficient of diode efficiency}

\begin{figure}[h]
\begin{center}
\includegraphics[width=0.48\textwidth]{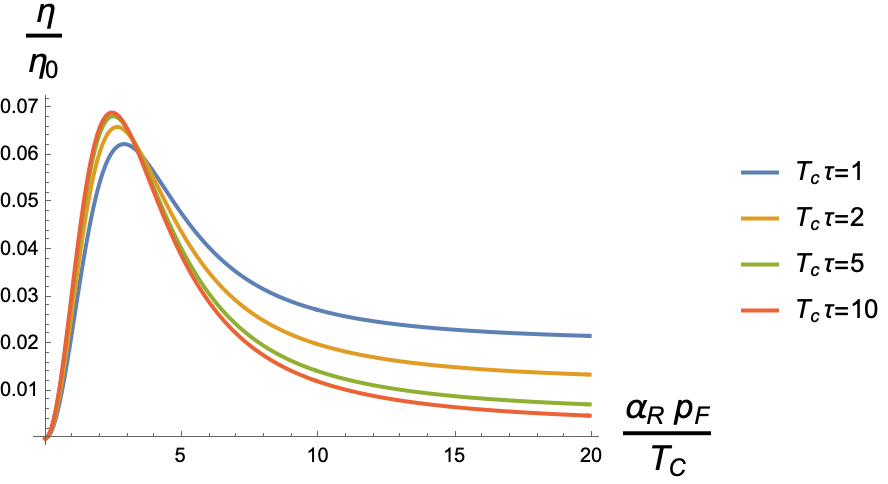} 
\includegraphics[width=0.48\textwidth]{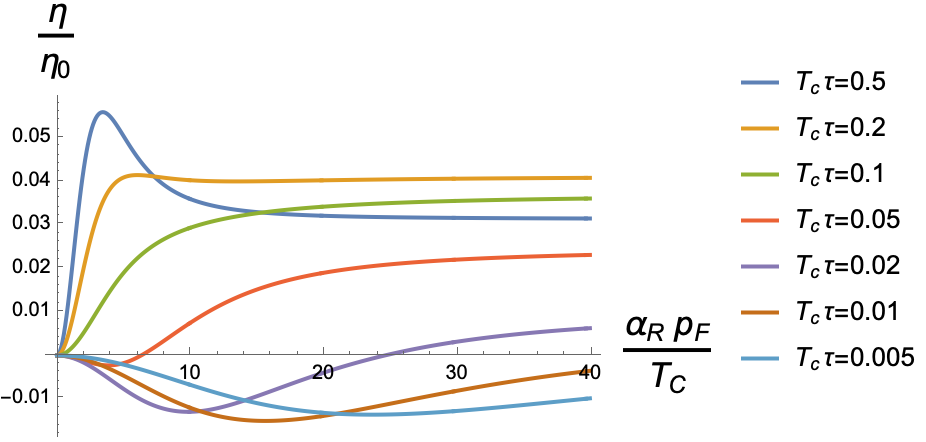} 
\caption{Normalized diode efficiency coefficient in the unit of $\eta_0=\frac{h}{T_c} \frac{\alpha_R}{v_F} \sqrt{t}$ plotted as a function of the SOC for different values of disorder strength that can be quantified by the dimensionless product $T_c\tau$ shown on the plot legends. The top panel corresponds to the ballistic limit and the bottom panel describes the diffusive regime.} 
\label{arbSOC1}
\end{center}
\end{figure}

In the limit of arbitrary $\frac{\alpha_R p_F}{T_c}$ ratio, the diode efficiency $\eta$ has the following expression,
\begin{equation}
\eta= \frac{h}{T_c} \frac{\alpha_R}{v_F} \mathcal{F}\left(\frac{1}{T_c \tau}, \frac{\alpha_R p_F}{T_c}\right) \sqrt{t}.
\end{equation}
He we introduced the reduced temperature $t=\frac{T_c-T}{T_c}$. The analytic form of the function $\frac{\eta}{\eta_0}=\mathcal{F}$ is cumbersome, so it is plotted versus $\frac{\alpha_R p_F}{T_c}$ in Fig. \ref{arbSOC1} for different values of $T_c\tau$. We chose the normalization to $\eta_0=\frac{h}{T_c} \frac{\alpha_R}{v_F} \sqrt{t}$.

We observe that \(\eta\) exhibits nonmonotonic behavior as a function of Rashba SOC strength and depends sensitively on disorder strength. As expected, \(\eta\) vanishes in the absence of SOC and generally asymptotes to a constant value as  \(\alpha_Rp_F\) increases.

At weak disorder (large \(T_c\tau\)), $\eta$ initially increases with 
\(\alpha_Rp_F\), reaches a maximum at $\alpha_Rp_F\sim T_c$, where interband pairing is strongest \cite{HasanShafferKhodasLevchenko24}, and then approaches the asymptote from above, as shown in Fig. \ref{arbSOC1}. In contrast, at strong disorder (small \(T_c\tau\)), $\eta$ initially decreases, becomes negative, reaches a local minimum, and then increases toward a positive asymptote, changing sign in the process. Analytically, we find that this transition occurs at 
\(T_c\tau\simeq 0.018\).

\begin{figure}[t!]
\begin{center}
\includegraphics[width=0.48\textwidth]{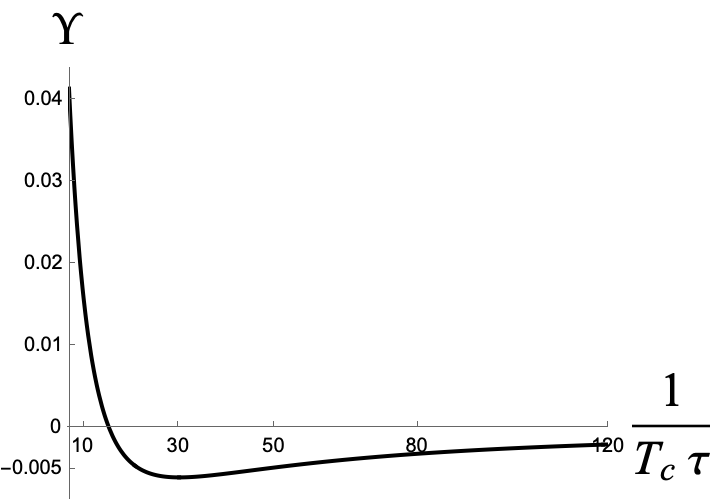} 
\includegraphics[width=0.48\textwidth]{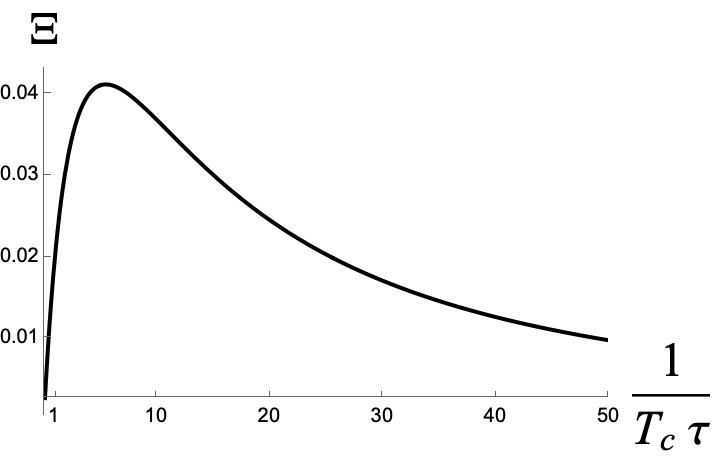} 
\caption{Plots for the dimensionless functions $\Upsilon$ and $\Xi$ versus the parameter $\frac{1}{T_c \tau}$ as introduced in Eqs. \eqref{eq:eta-weakSOC} and \eqref{eq:infSOC} respectively. }  \label{fig:Upsilon-Xi}
\end{center}
\end{figure}

In the limit of weak Rashba SOC, i.e. $\varkappa=2\alpha_R p_F \tau \ll 1$ or, equivalently, $\alpha_R p_F \ll \tau^{-1}, T_c$, the diode efficiency reduces, to leading order:
\begin{equation}\label{eq:eta-weakSOC}
\eta= \frac{h}{T_c} \frac{\alpha_R}{v_F} \left(\frac{\alpha_R p_F}{T_c}\right)^2 \Upsilon\left(\frac{1}{T_c \tau}\right) \sqrt{t},
\end{equation}
The analytic form of the function $\Upsilon$ is cumbersome. It is plotted in Fig. \ref{fig:Upsilon-Xi} (top panel). Importantly, it changes sign around $\frac{1}{T_c\tau}\sim15$, at which point increasing disorder actually increases the diode effect.

A simple analytic expressions for $\eta$ can be found for the clean case $\left(T_c \gg \tau^{-1} \gg \alpha_R p_F\right)$ as:
\begin{equation}\label{eq:cleanweakSOC}
\eta= \left[0.043-\frac{0.019}{T_c \tau}\right] \frac{h}{T_c} \frac{\alpha_R}{v_F} \left(\frac{\alpha_R p_F}{T_c}\right)^2 \sqrt{t},
\end{equation}
and for the diffusive case $\left(\tau^{-1} \gg T_c \gg \alpha_R p_F\right)$ as:
\begin{equation}\label{eq:dirtyweakSOC}
\eta= -0.307\left(T_c \tau\right)^{\frac{3}{2}} \frac{h}{T_c} \frac{\alpha_R}{v_F} \left(\frac{\alpha_R p_F}{T_c}\right)^2 \sqrt{t}.
\end{equation}
In particular, \(\eta\) approaches zero from below as \((T_c\tau)^\frac{3}{2}\).

In the strong SOC limit $\varkappa=2\alpha_R p_F \tau \gg 1$ or, equivalently, $ \tau^{-1}, T_c\ll \alpha_R p_F \rightarrow\infty$, we find instead 
\begin{equation}\label{eq:infSOC}
\eta= \frac{h}{T_c} \frac{\alpha_R}{v_F} \Xi\left(\frac{1}{T_c \tau}\right) \sqrt{t},
\end{equation}
The dimensionless function $\Xi$ is plotted in Fig. \ref{fig:Upsilon-Xi}. It has the following asymptotic limits that can be extracted analytically 
in the weak disorder limit $\left(\alpha_R p_F \gg T_C \gg \tau^{-1}\right)$:
\begin{equation}\label{eq:cleaninfSOC}
\eta= \frac{0.027}{T_c \tau} \frac{h}{T_c} \frac{\alpha}{v_F} \sqrt{t},
\end{equation}
and for the diffusive case $\left(\alpha_R p_F \gg \tau^{-1} \gg T_c\right)$:
\begin{equation}\label{eq:dirtyinfSOC}
\eta= \left[7.44+0.47\ln\left(T_c \tau\right)\right]\left(T_c \tau\right)^{\frac{3}{2}} \frac{h}{T_c} \frac{\alpha}{v_F} \sqrt{t}.
\end{equation}
It is important to note here that the diode efficiency $\eta$ becomes independent of the parameter $\frac{\alpha_R p_F}{T_C}$ in the strong SOC limit.

\begin{figure}[h]
\begin{center}
\includegraphics[width=0.48\textwidth]{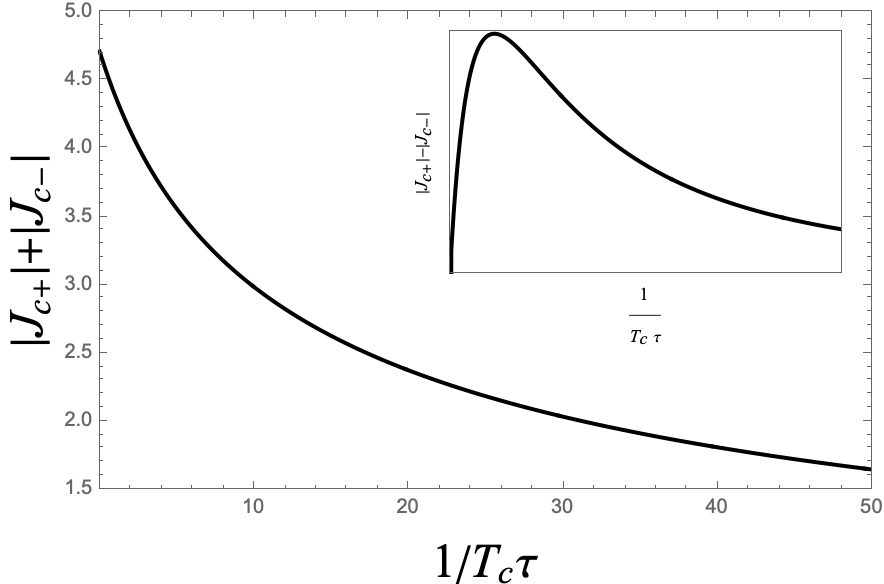} 
\caption{Dependence of the critical currents on the strength of disorder. The main plot represents the sum of the magnitude of the nonreciprocal critical currents $|J_{c+}|+|J_{c-}|$ versus $\frac{1}{T_c\tau}$ shown in the unit of $v_F T_c t^{\frac{3}{2}}$. The inset shows the difference $|J_{c+}|-|J_{c-}|$ plotted versus $\frac{1}{T_c\tau}$ in the unit of $h \alpha_R t^2$.
}  \label{jmagsuminf}
\end{center}
\end{figure}

As we can see from Fig. \ref{fig:Upsilon-Xi}, diode efficiency $\eta$ increases with disorder strength, reaches a maximum and then vanishes in the dirty limit, $T_c \tau \ll 1$ as Eq. (\ref{eq:dirtyinfSOC}). In particular, \(\eta=0\) in the clean limit, as shown in \cite{IlicBergeret22}, which is a result of an accidental approximate symmetry in this limit \cite{HaimLevchenkoKhodas22}. Disorder breaks this symmetry, as we discuss below, and as a result induces a nonvanishing diode effect at weak disorder. Note that although \(\eta\) increases with disorder, the critical currents $|J_{c+}|$ and $|J_{c-}|$ both decrease monotonically with increasing disorder as expected, see Fig. \ref{jmagsuminf}, but they decrease at different rates such that the difference \(|J_{c+}|-|J_{c-}|\) increases initially. With stronger disorder, $\Xi$ reaches a maximum and begins to decrease.

\subsection{Intraband Limit in Helical Basis}

As has been shown in \cite{IlicBergeret22}, in the clean case the SDE vanishes in the limit of strong Rashba SOC. This occurs due to an approximate symmetry that holds to linear order in \(h\) in this case \cite{HasanShafferKhodasLevchenko24}. A surprising result that follows from Eq. (\ref{eq:cleaninfSOC}) is that \(\eta\) becomes nonzero in the presence of weak disorder and grows with increasing disorder, implying a disorder-driven SDE. As we now show, this happens due to singlet/triplet pairing mixing induced by the disorder \cite{edelstein_ginzburg-landau_2021}. Here we work in the helical basis and drop the interband pairing terms. The approximate symmetry that guarantees the vanishing of \(\eta\) follows from the fact that both the \(\alpha(q)\) and \(\beta(q)\) had the following Taylor expansions to linear order in \(h\) (for simplicity we take \(\mathbf{q}\perp\mathbf{h})\):
\begin{align}
    \alpha(q)&\approx\sum_{n} A_n(v_F q)^{2n-1}[4n\Delta\nu h +\nu v_F q]\\
    \beta(q)&\approx\sum_{n} B_n(v_F q)^{2n-1}[4n\Delta\nu h +\nu v_F q]\,,
\end{align}
The symmetry holds as long as \(\alpha_{2n-1}/\alpha_{2n}=\beta_{2n-1}/\beta_{2n}\propto n\). Following the calculation in \cite{AttiasKhodas24}, we find that to leading order in \(\tau^{-1}\)
\begin{widetext}
\begin{align}\label{eq:alphaIntraband}
\alpha(q)&=\alpha_{\infty}(q)-\frac{\pi}{8\tau T}\sum_\lambda\nu_\lambda\int\text{sech}^2\left(\frac{(qv_F+\lambda h)\cos\theta}{2T}\right)\frac{d\theta}{2\pi}+\\
&+\frac{1}{4\tau}\sum_{\lambda\lambda'}\frac{\pi\nu_\lambda\nu_{\lambda'}}{\nu(2\pi)^2}\iint(1+\lambda\lambda'\cos\theta\cos\theta')\frac{\tanh\left(\frac{qv_F+\lambda h}{2T}\cos\theta\right)-\tanh\left(\frac{qv_F+\lambda' h}{2T}\cos\theta'\right)}{(qv_F+\lambda h)\cos\theta-(qv_F+\lambda' h)\cos\theta'}d\theta\, d\theta'\nonumber
\end{align}
\end{widetext}
where \(\alpha_{\infty}(q)\) is the expression in the clean limit \(\tau\rightarrow\infty\); here \(\theta\) is the angle along the Fermi surface. The first correction on the first line in Eq. (\ref{eq:alphaIntraband}) is due to the finite lifetime of the electrons in presence of disorder and can be shown to respect the accidental symmetry to linear order in \(h\). The term on the second line originates from the vertex correction due to disorder induced singlet/triplet mixing, and this term violates the approximate symmetry as the ratio \(\alpha_{2n-1}/\alpha_{2n}\) is no longer linear in \(n\) (see Appendix \ref{AppD} for full expression). We therefore attribute the SDE in the intraband limit to the disorder-induced singlet-triplet mixing.

\section{Summary}\label{SecV}

In this work, using a diagrammatic approach to derive Ginzburg-Landau functional, we studied the superconducting diode coefficient $\eta$ in a disordered Rashba superconductor with an in-plane magnetic field and found that disorder has a nontrivial effect on the SDE. While disorder is generally detrimental to superconductivity, our key finding is that in certain regimes, weak or moderate disorder can enhance or even induce the SDE.

For weak Rashba SOC, we observed that disorder initially decreases 
$\eta$ but eventually leads to a sign change, increasing $|\eta|$ for 
 \(1/(T_c\tau)<30\). Notably, this sign reversal is unrelated to the transition between the so-called weak and strong helical phases, which also induces a sign change in the diode effect \cite{IlicBergeret22, IkedaDaidoYanase22}. In the strong SOC limit, where the SDE vanishes due to an approximate symmetry, the effect is even more striking: weak disorder induces the SDE by mixing singlet and triplet order parameters.

Our results highlight the potential importance of disorder effects in SDE studies, particularly in the context of sign-changing $\eta$ observed in some Josephson diode effect experiments \cite{MarginedaGiazotto23, LotfizadehShabani24}. It would also be interesting to extend this study to other types of disorder and SOC, as well as to explore disorder effects in other forms of nonreciprocal superconducting transport, such as the JDE and AJE, which have received even less attention than the SDE \cite{LidalDanon23}. We leave these questions for future work.

\section*{Acknowledgments}

We thank Stefan Ilic for useful discussions. This work was financially supported by the National Science Foundation (NSF), Quantum Leap Challenge Institute for Hybrid Quantum Architectures and Networks Grant No. OMA-2016136 (D. S.). 
The work of J. H. was supported by the NSF Grant No. DMR-2203411. 
M. K. acknowledges financial support from the Israel Science Foundation, Grant No. 2665/20. A. L. gratefully acknowledges NSF Grant No. DMR-2452658 and H. I. Romnes Faculty Fellowship provided by the University of Wisconsin-Madison Office of the Vice Chancellor for Research and Graduate Education with funding from the Wisconsin Alumni Research Foundation.


\appendix

\begin{figure}
\includegraphics[width=0.48\textwidth]{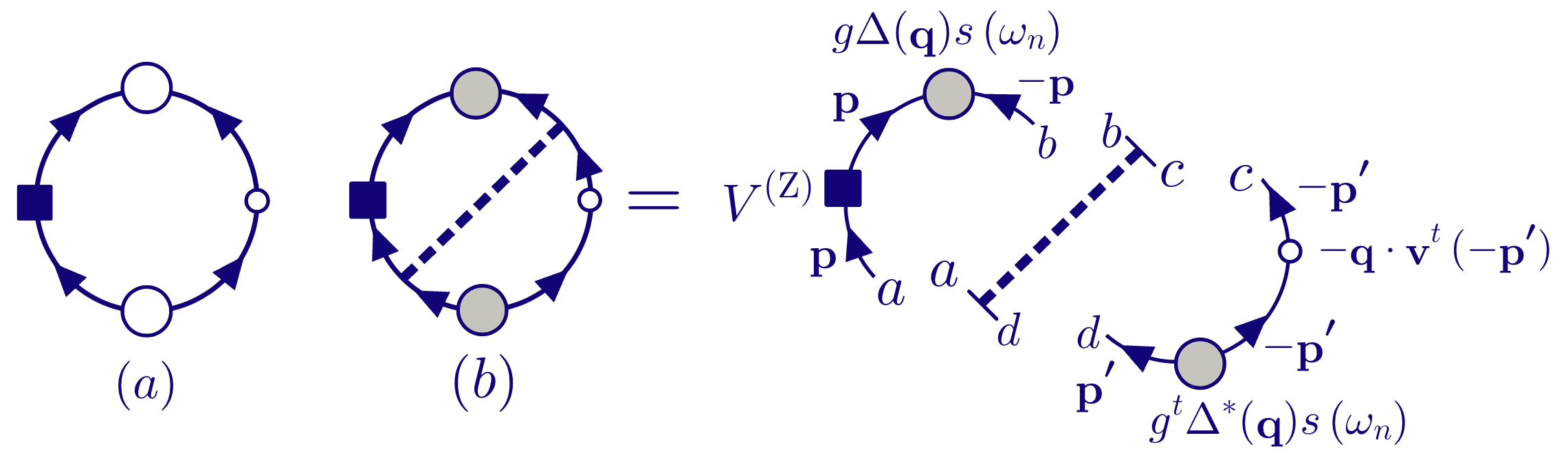} 
\caption{One of the skeleton diagrams responsible for the anomalous GL coefficient $\alpha_1$ in the clean limit is presented in $(a)$ and $(b)$ represents the same diagram after the insertion of impurity ladder and impurity renormalized order parameter in the disordered limit. The split form of this diagram is shown on the right.}  \label{fig12}
\end{figure}

\section{Evaluation of the diagram in Fig. \ref{fig12}$(a)$}\label{AppA}

In this appendix, we evaluate one of the diagrams responsible for the anomalous GL coefficient $\alpha_1$ shown in Fig. \ref{fig12}$(a)$ in a clean superconductor to illustrate that this diagram is purely diagonal in the helical basis in the limit $\kappa\gg1$, i.e. when the SOC energy scale characterized by $\alpha_R p_F$ is much larger than the critical temperature $T_c$. This diagram can be decomposed into a product of two factors, the first of which, known as the slow factor stemming from the integration over variables which vary slowly in coordinate space (magnetic field $\mathbf{B}(\mathbf{q^{\prime}})$, and the gap function $\Delta(\mathbf{q})$) is given by
\begin{equation}
S_{\mathrm{\ref{fig12}(a)}}=\int_{\mathbf{q}} h_m |\Delta\left(\mathbf{q}\right)|^2 q_j,
\end{equation}
where $h_m=\mu_B B_m$ is the $m$th component of $\mathbf{h}$. 
Its quick factor $Q_{\mathrm{\ref{fig12}(a)}}$ is given by
\begin{equation}
\begin{aligned}
Q_{\mathrm{\ref{fig12}(a)}}= & T \sum_{\omega_n} \int_{\mathbf{p}} \operatorname{Tr}\{\mathbf{\sigma}_m G\left(i \omega_n, \mathbf{p}\right)
G^{(\mathrm{r})}\left(i \omega_n, \mathbf{p}\right) \\& \times v_j(\mathbf{p}) G^{(\mathrm{r})}\left(i \omega_n, \mathbf{p}\right) G\left(i \omega_n, \mathbf{p}\right)\}\\
=& T \sum_{\omega_n} \int_{\mathbf{p}} \sum_{\lambda \lambda^{\prime}} \operatorname{Tr}\left\{\mathbf{\sigma}_m \Pi^{(\lambda)}(\mathbf{p}) v_j(\mathbf{p}) \Pi^{(\lambda^{\prime})}(\mathbf{p})\right\}\\
&\times G_{(\lambda)} G_{(\lambda)}^{\text {(r)}} G_{(\lambda^{\prime})} G_{(\lambda^{\prime})}^{(\text{r})},
\end{aligned}
\end{equation}
where $\int_{\mathbf{p}}=2\pi\nu \int \frac{d \xi}{2 \pi} \int \frac{d \hat{p}}{2 \pi}$ and $G^{(0)}\left(i \omega_n, \mathbf{p}\right)$ has been replaced by $G\left(i \omega_n, \mathbf{p}\right)$ for notational convenience. Here, the angle integral has the following expression,
\begin{equation}
\begin{aligned}
&\int \frac{d \hat{p}}{2 \pi} \operatorname{Tr}\left\{\mathbf{\sigma}_m \Pi^{(\lambda)}(\mathbf{p}) v_j(\mathbf{p}) \Pi^{(\lambda^{\prime})}(\mathbf{p})\right\} \\
& =\frac{1}{2} \varepsilon_{m j i} c_i\left[\frac{1}{2}(\lambda+\lambda^{\prime}) \frac{p}{m}+\alpha_R\right],
\end{aligned}
\end{equation}
Using this expression $Q_{\mathrm{\ref{fig12}(a)}}$ takes the following form 
\begin{equation}\label{eq:Q11a}
\begin{aligned}
&Q_{\mathrm{\ref{fig12}(a)}} = \pi \nu \varepsilon_{m j i} c_i T \sum_{\omega_n} \sum_{\lambda=\pm1} \int \frac{d \xi}{2 \pi}\\&  \biggl\{\left(\lambda \frac{p}{m}+\alpha_R\right)G_{(\lambda)}^2 \left[G_{(\lambda)}^{\text {(r)}}\right]^2+\alpha_R G_{(\lambda)} G_{(\lambda)}^{\text {(r)}} G_{(-\lambda)} G_{(-\lambda)}^{(\text{r})}\biggr\},
\end{aligned}
\end{equation}
where the last term in Eq. (\ref{eq:Q11a}) is not diagonal in the helical indices, i.e. $\lambda=-\lambda^{\prime}$ terms also contribute to $Q_{\mathrm{\ref{fig12}(a)}}$. This can be analyzed and simplified using the following integrals:
\begin{subequations}
\begin{equation}
\int \frac{d \xi}{2 \pi} G_{(\lambda)}^2 \left[G_{(\lambda)}^{(r)}\right]^2=(1-\lambda \delta) \frac{2}{\left[2\omega_n\right]^3},
\end{equation}
\begin{equation}
\int \frac{d \xi}{2 \pi} \left(\frac{p}{m}\right) G_{(\lambda)}^2 \left[G_{(\lambda)}^{(r)}\right]^2=(1-2\lambda \delta) \frac{2}{\left[2\omega_n\right]^3},
\end{equation}
\begin{equation}\label{eq:off_diag22}
\int \frac{d \xi}{2 \pi} G_{(\lambda)} G_{(\lambda)}^{(r)} G_{(-\lambda)} G_{(-\lambda)}^{(r)}=\frac{1}{\left|\omega_n\right|} \cdot \frac{1}{\left[2\left|\omega_n\right|\right]^2+(2\alpha_R p_F)^2}.
\end{equation}
\end{subequations}
We note that the integral in Eq. (\ref{eq:off_diag22}) vanishes in the limit $\kappa\to\infty$. Thus the off-diagonal contribution to $Q_{\mathrm{\ref{fig12}(a)}}$ vanishes giving us a purely diagonal form of $Q_{\mathrm{\ref{fig12}(a)}}$ in terms of helical indices in the limit of strong SOC.


\section{Evaluation of the diagram in Fig. \ref{fig12}$(b)$}\label{AppB}

In this appendix, we evaluate one of the diagrams responsible for the anomalous GL coefficient $\alpha_1$ shown in Fig. \ref{fig12}$(b)$ in a disordered superconductor to illustrate that the impurity ladder mixes the two helical bands. 

In analogy to the previous example, the slow factor of this diagram is given by
\begin{equation}
S_{\mathrm{\ref{fig12}(b)}}=\int_{\mathbf{q}} h_m |\Delta\left(\mathbf{q}\right)|^2 q_j,
\end{equation}
Its quick factor $Q_{\mathrm{\ref{fig12}(b)}}$ is given by
\begin{equation}
\begin{aligned}
Q_{\mathrm{\ref{fig12}(b)}}= & \left(-n_{\text {imp }}|U|^2\right) T \sum_{\omega_n} s^2\left(\omega_n\right) \\
& \times \operatorname{Tr}\left\{L\left(\omega_n\right) \circ T\left(\omega_n\right) \circ R\left(\omega_n\right)\right\},
\end{aligned}
\end{equation}
where $T\left(\omega_n\right)$ is given by Eq. (\ref{eq:Tmatrix}), $L\left(\omega_n\right)$ and $R\left(\omega_n\right)$ denote the left and right fragments of the diagram in its split form (see Fig. \ref{fig12}$(b)$): 
\begin{equation}
\begin{aligned}
L_{ab}&=\int_{\mathbf{p}}\left\{G\left(i \varpi_n, \mathbf{p}\right) \mathbf{\sigma}_m G\left(i \varpi_n, \mathbf{p}\right) G^{(\mathrm{r})}\left(i\varpi_n, \mathbf{p}\right)\right\}_{\rho \beta}\\
&=\int_{\mathbf{p}} \sum_{\lambda \lambda^{\prime}} G_{(\lambda^{\prime})} G_{(\lambda)} G_{(\lambda)}^{(\mathrm{r})}\left\{\Pi^{(\lambda^{\prime})} \sigma_m \Pi^{(\lambda)}\right\}_{ab}.
\end{aligned}
\end{equation}
Here, the angular integral has the following expression,
\begin{equation}
\int \frac{d \hat{p}}{2 \pi}\left[\Pi^{(\lambda^{\prime})} \sigma^m \Pi^{(\lambda)}\right]_{ab}=\frac{1}{4}\left[\sigma_m-\lambda \lambda^{\prime} c_m(\mathbf{c} \cdot \bm{\sigma})\right]_{ab}
\end{equation}
Using this expression $L_{ab}$ takes the following form
\begin{equation}\label{eq:Lb}
\begin{aligned}
L_{ab}=&2\pi\nu\sum_{\lambda= \pm 1} \left\{\frac{1}{4}\left[\sigma_m-c_m(\mathbf{c} \cdot \bm{\sigma})\right]_{ab} \int \frac{d \xi}{2 \pi} G_{(\lambda)}^2 G_{(\lambda)}^{(r)}\right. \\
& \left.+\frac{1}{4}\left[\sigma_m+c_m(\mathbf{c} \cdot \bm{\sigma})\right]_{ab} \int \frac{d \xi}{2 \pi} G_{(\lambda)} G_{(\lambda)}^{(r)} G_{(-\lambda)}\right\},
\end{aligned}
\end{equation}
where the last term in Eq. (\ref{eq:Lb}) is not diagonal in the helical indices, i.e. $\lambda=-\lambda^{\prime}$ terms also contribute to $L_{ab}$. This can be simplified using the following integrals:
\begin{subequations}
\begin{equation}
\int \frac{d \xi}{2 \pi} G_{(\lambda)}^2 G_{(\lambda)}^{(r)}=(1-\lambda \delta) \frac{i \operatorname{sgn}\left(\omega_n\right)}{\left[2 \varpi_n\right]^2},
\end{equation}
\begin{equation}\label{eq:off_diag21}
\int \frac{d \xi}{2 \pi} G_{(\lambda)} G_{(\lambda)}^{(r)} G_{(-\lambda)}=\frac{i \operatorname{sgn}\left(\omega_n\right)}{\left(2\varpi_n\right)^2-4 i \lambda\left|\varpi_n\right| \alpha_R p_F}.
\end{equation}
\end{subequations}
We note that the integral in Eq. (\ref{eq:off_diag21}) vanishes in the limit $\kappa\to\infty$, giving us a purely diagonal left fragment of the diagram $L_{\rho \beta}$ in the limit of strong SOC. We finally get for $L_{ab}$:
\begin{equation}\label{eq:left}
L_{ab}=\pi\nu i \operatorname{sgn}\omega_n\left[\frac{\sigma_m-c_m(\mathbf{c} \cdot \bm{\sigma})}{\left[2\varpi_n\right]^2}+\frac{\sigma_m+c_m(\mathbf{c} \cdot \bm{\sigma})}{\left(2\varpi_n\right)^2+(2\alpha_Rp_F)^2}\right]_{ab},
\end{equation}

We proceed to evaluate the right fragment of the diagram,
\begin{equation}
\begin{aligned}
R_{cd}&=\int_{\mathbf{p}}\left\{G^{(r)}\left(i\varpi_n, \mathbf{p}\right) v^j(\mathbf{p}) G^{(r)}\left(i\varpi_n, \mathbf{p}\right) G\left(i\varpi_n, \mathbf{p}\right)\right\}_{cd}\\
&=\int_{\mathbf{p}} \sum_{\lambda \lambda^{\prime}} G_{(\lambda^{\prime})}^{(r)} G_{(\lambda)}^{(r)} G_{(\lambda)}\left\{\Pi^{(\lambda^{\prime})} v^j(\mathbf{p}) \Pi^{(\lambda)}\right\}_{cd},
\end{aligned}
\end{equation}
where the angle integral is given by
\begin{equation}
\begin{aligned}
&\int \frac{d \hat{p}}{2 \pi}\left[\Pi^{(\lambda^{\prime})} v^j(\mathbf{p}) \Pi^{(\lambda)}\right]_{cd}\\&=\frac{1}{4}(\mathbf{c} \times \bm{\sigma})^j_{cd}\left[\alpha_R+\frac{1}{2}(\lambda+\lambda^{\prime})\left(\frac{p}{m}\right)\right]
\end{aligned}
\end{equation}
Using this expression $R_{ab}$ takes the form
\begin{equation}\label{eq:Rb}
\begin{aligned}
R_{cd}=&\pi\nu \frac{1}{2}(\mathbf{c} \times \bm{\sigma})^j_{cd}\sum_{\lambda= \pm 1} \int \frac{d \xi}{2 \pi}\biggl\{\left[\alpha_R +\lambda\left(\frac{p}{m}\right)\right]\\& \times G_{(\lambda)} \left[G_{(\lambda)}^{(r)}\right]^2
+\alpha_R G_{(\lambda)} G_{(\lambda)}^{(r)} G_{(-\lambda)}^{(r)}\biggr\}
\end{aligned}
\end{equation}
where the last term in Eq. (\ref{eq:Rb}) is not diagonal in the helical indices, i.e. $\lambda=-\lambda^{\prime}$ terms also contribute to $R_{ab}$. This can be simplified using the following integrals:
\begin{subequations}
\begin{equation}
\int \frac{d \xi}{2 \pi} G_{(\lambda)} \left[G_{(\lambda)}^{(r)}\right]^2=(1-\lambda \delta) \frac{i \operatorname{sgn}\left(\omega_n\right)}{\left[2 \varpi_n\right]^2},
\end{equation}
\begin{equation}\label{eq:off_diag21r}
\int \frac{d \xi}{2 \pi} G_{(\lambda)} G_{(\lambda)}^{(r)} G_{(-\lambda)}^{(r)}=\frac{i \operatorname{sgn}\left(\omega_n\right)}{\left(2\varpi_n\right)^2+4 i \lambda\left|\varpi_n\right| \alpha_R p_F}.
\end{equation}
\end{subequations}
We note that the integral in Eq. (\ref{eq:off_diag21r}) vanishes in the limit $\kappa\to\infty$, giving us a purely diagonal right fragment of the diagram $R_{cd}$ in the limit of strong SOC. We finally get for $R_{cd}$:
\begin{equation}\label{eq:right}
R_{cd}=-\pi\nu i \operatorname{sgn}\left(\omega_n\right)\left[\frac{\alpha_R(\mathbf{c} \times \bm{\sigma})_j (2\alpha_Rp_F)^2}{\left[2\varpi_n\right]^2\left[\left(2 \varpi_n\right)^2+(2\alpha_Rp_F)^2\right]}\right]_{cd}.
\end{equation}
From the form of Eq. (\ref{eq:Tmatrix}), (\ref{eq:left}) and (\ref{eq:right}), it can be seen that $T\left(\omega_n\right)$ connects $L\left(\omega_n\right)$ and $R\left(\omega_n\right)$ with its last term
\begin{equation}
\frac{1}{2} \frac{(\mathbf{c} \times \bm{\sigma})_{ba}^l(\mathbf{c} \times \bm{\sigma})_{dc}^l}{1-v\left(\omega_n\right)}.
\end{equation}
We finally get for $Q_{\mathrm{\ref{fig12}(b)}}$:
\begin{equation}\label{eq:Q11b}
\begin{aligned}
Q_{\mathrm{\ref{fig12}(b)}}=&-\frac{1}{4\pi\nu\tau} T \sum_{\omega_n} s^2\left(\omega_n\right) \frac{1}{1-v\left(\omega_n\right)} \\
& \times \operatorname{Tr}\left[L\left(\omega_n\right)(\mathbf{c} \times \bm{\sigma})^l\right] \operatorname{Tr}\left[R\left(\omega_n\right)(\mathbf{c} \times \bm{\sigma})^l\right]
\end{aligned}
\end{equation}
We note from the Eq. (\ref{eq:Q11b}) that $Q_{\mathrm{\ref{fig12}(b)}}$ is not diagonal in the helical indices even though the left and right fragments of the diagram $L\left(\omega_n\right)$ and $R\left(\omega_n\right)$ are individually diagonal in the helical indices in the limit of strong SOC. The mere presence of the impurity ladder mixes the two helical bands, even in the limit of strong SOC. Computing the traces, we finally get from Eq. (\ref{eq:Q11b}):
\begin{equation}
\begin{aligned}
Q_{\mathrm{\ref{fig12}(b)}}&=\varepsilon_{j m i} c^i Q_{(l)} ; \text { where } \\
Q_{(l)}&=\nu\alpha_R \sum_{\omega_n>0} \frac{\pi T}{\omega_n^2} \cdot \frac{\tau \varkappa^2\left[2\left(2 \tau\omega_n+1\right)^2+\varkappa^2\right]}{\left(2 \tau\omega_n+1\right)\left[\left(2 \tau\omega_n+1\right)^2+\varkappa^2\right]}\\
& \cdot \frac{1}{4 \tau\omega_n\left(2 \tau\omega_n+1\right)^2+\varkappa^2\left(4 \tau\omega_n+1\right)}.
\end{aligned}
\end{equation}

\section{GL coefficients in terms of digamma functions}\label{AppC}

Completing the Matsubara summation and writing the GL coefficients in terms of the digamma function $\psi(x)$ and its $N^{\text{th}}$ order derivatives, 
\begin{equation}
\psi^{(N)}(z)=(-1)^{N+1}N!\sum^{\infty}_{n=0}\frac{1}{(n+z)^{N+1}},
\end{equation}
as well as Riemann zeta-function,  
\begin{equation}
\zeta(s)=\sum^{\infty}_{n=1}\frac{1}{n^s},
\end{equation}
we get for the conventional coefficients:
\begin{widetext}
\begin{subequations}
\begin{align}
& \alpha_0=-\nu\left(\frac{T_c-T}{T_c}\right), \qquad 
\alpha_2=\frac{\nu v_F^2 \tau}{16 T}\left[\pi-8 T \tau\left(\psi\left(\frac{1}{2}+\frac{1}{4 \pi T \tau}\right)-\psi\left(\frac{1}{2}\right)\right)\right], \qquad
\beta_0=\frac{7\zeta(3) \nu}{16 \left(\pi T\right)^2},\\
& \alpha_4=-\frac{\nu v_F^4 \tau^2}{256 \left(\pi T\right)^2}\left[28 \zeta(3)-12 \pi^3 T \tau+96 \left(\pi T \tau\right)^2 \left(\psi\left(\frac{1}{2}+\frac{1}{4 \pi T \tau}\right)-\psi\left(\frac{1}{2}\right)\right)-\psi^{(2)}\left(\frac{1}{2}+\frac{1}{4 \pi T \tau}\right)\right],\\
& \beta_2=-\frac{\nu v_F^2 \tau}{192 \left(\pi T\right)^3}\left[84 \pi T \tau \zeta(3)-\pi^4+48 \left(\pi T \tau\right)^2 \psi^{(1)}\left(\frac{1}{2}+\frac{1}{4 \pi T \tau}\right)-192\left(\pi T \tau\right)^3 \left(\psi\left(\frac{1}{2}+\frac{1}{4 \pi T \tau}\right)-\psi\left(\frac{1}{2}\right)\right)\right].
\end{align}
\end{subequations}

In the limit of weak Rashba SOC, i.e. $\varkappa\ll 1$ or, equivalently, $\alpha_R p_F \ll \tau^{-1}, T_c$, we get for the anomalous coefficients:
\begin{subequations}
\begin{align}
\alpha_1=&-\frac{\nu\alpha_R (\left[\mathbf{h} \times \hat{\mathbf{q}}\right] \cdot \mathbf{c}) \left(\alpha_R p_F \tau\right)^2}{2\left(\pi T\right)^2}\left[7 \zeta(3)-4 \pi^3 T \tau+48 \left(\pi T \tau\right)^2 \left(\psi\left(\frac{1}{2}+\frac{1}{4 \pi T \tau}\right)-\psi\left(\frac{1}{2}\right)\right)\right.\nonumber\\& \left.-4 \pi T \tau\psi^{(1)}\left(\frac{1}{2}+\frac{1}{4 \pi T \tau}\right)\right],\\
\alpha_3=&\frac{\nu\alpha_R v_F^2 (\left[\mathbf{h} \times \hat{\mathbf{q}}\right] \cdot \mathbf{c})\left(\alpha_R p_F \tau\right)^2}{3072\left(\pi T\right)^4}\left[96\pi^5 T\tau+3840\pi^5\left(T\tau\right)^3-26880\left(\pi T\tau\right)^2 \zeta(3)-30720 \left(\pi T \tau\right)^3\psi^{(1)}\left(\frac{1}{2}+\frac{1}{4 \pi T \tau}\right) \right.\nonumber\\
& \left.+92160 \left(\pi T \tau\right)^4 \left(\psi\left(\frac{1}{2}+\frac{1}{4 \pi T \tau}\right)-\psi\left(\frac{1}{2}\right)\right)+2880\left(\pi T \tau\right)^2\psi^{(2)}\left(\frac{1}{2}+\frac{1}{4 \pi T \tau}\right)-96\pi T \tau \psi^{(3)}\left(\frac{1}{2}+\frac{1}{4 \pi T \tau}\right)\right],\\
\beta_1=&\frac{\nu\alpha_R (\left[\mathbf{h} \times \hat{\mathbf{q}}\right] \cdot \mathbf{c})\left(\alpha_R p_F \tau\right)^2}{24\left(\pi T\right)^4}\left[2184\left(\pi T\tau\right)^2 \zeta(3)+186\zeta(5)-8\pi^5 T\tau-960\pi^5\left(T\tau\right)^3-960 \left(\pi T \tau\right)^3\psi^{(1)}\left(\frac{1}{2}+\frac{1}{4 \pi T \tau}\right) \right.\nonumber\\
& \left.+11520 \left(\pi T \tau\right)^4 \left(\psi\left(\frac{1}{2}+\frac{1}{4 \pi T \tau}\right)-\psi\left(\frac{1}{2}\right)\right)+36\left(\pi T \tau\right)^2\psi^{(2)}\left(\frac{1}{2}+\frac{1}{4 \pi T \tau}\right)-\pi T \tau \psi^{(3)}\left(\frac{1}{2}+\frac{1}{4 \pi T \tau}\right)\right],
\end{align}
\end{subequations}

In the strong SOC limit $\varkappa \gg 1$ or, equivalently, $ \tau^{-1}, T_c\ll \alpha_R p_F$, we find:
\begin{subequations}
\begin{align}
\alpha_1=&-\frac{\nu\alpha_R \tau (\left[\mathbf{h} \times \hat{\mathbf{q}}\right] \cdot \mathbf{c})}{2T}\left[\pi-16 T \tau \left(\psi\left(\frac{1}{2}+\frac{1}{8 \pi T \tau}\right)-\psi\left(\frac{1}{2}\right)\right)\right],\\
\alpha_3=&\frac{\nu\alpha_R v_F^2 \tau^2 (\left[\mathbf{h} \times \hat{\mathbf{q}}\right] \cdot \mathbf{c})}{64\left(\pi T\right)^2}\left[112\zeta(3)-44\pi^3 T\tau +160 \left(\pi T \tau\right)^2\left(\psi\left(\frac{1}{2}+\frac{1}{4 \pi T \tau}\right)-\psi\left(\frac{1}{2}\right)\right)+3 \psi^{(2)}\left(\frac{1}{2}+\frac{1}{4 \pi T \tau}\right) \right.\nonumber\\
& \left.-512 \left(\pi T \tau\right)^2\left(\psi\left(\frac{1}{2}+\frac{1}{8 \pi T \tau}\right)-\psi\left(\frac{1}{2}\right)\right)-80\pi T \tau\psi^{(1)}\left(\frac{1}{2}+\frac{1}{4 \pi T \tau}\right) +192\pi T \tau\psi^{(1)}\left(\frac{1}{2}+\frac{1}{8 \pi T \tau}\right)\right],\\
\beta_1=&\frac{\nu\alpha_R \tau(\left[\mathbf{h} \times \hat{\mathbf{q}}\right] \cdot \mathbf{c})}{24\left(\pi T\right)^3}\left[\pi^4 -168\pi T\tau \zeta(3) +1536 \left(\pi T \tau\right)^3\left(\psi\left(\frac{1}{2}+\frac{1}{8 \pi T \tau}\right)-\psi\left(\frac{1}{2}\right)\right)\right. \nonumber \\
& \left. -192\left(\pi T \tau\right)^2 \psi^{(1)}\left(\frac{1}{2}+\frac{1}{8 \pi T \tau}\right)\right],
\end{align}
\end{subequations}
\end{widetext}
The general analytical expressions for the anomalous coefficients are cumbersome and are presented here only for the above two different limiting cases.

\section{Calculation of \(\alpha(q)\) in the helical basis in the strong SOC limit}\label{AppD}

The calculation for \(\alpha(q)\) with short-range disorder for the intraband case in the helical basis (with Rashba SOC and s-wave pairing interactions) has been carried out in \cite{AttiasKhodas24}, here we follow their formalism with slight modifications. We have
\begin{widetext}
\[\alpha(q)=-V^{-1}/2-\frac{T}{4}\sum_n\int_{0}^{2\pi}\left(\Gamma_+(\theta)\Pi_{++}(\theta,\mathbf{q},\omega_n)-\Gamma_-(\theta)\Pi_{--}(\theta,\mathbf{q},\omega_n)\right)\frac{d\theta}{2\pi}\]
\end{widetext}
where
\[\Pi_{\lambda\lambda}=\frac{2\pi \nu_\lambda}{2|\omega_n|+\tau^{-1}+2i \mathbf{v}_F\cdot\mathbf{Q}_\lambda\text{sgn}[\omega_n]}\]
with \(\tau^{-1}=2\pi\nu n_{\text{imp}}|U|^2\), \(\nu_\lambda=\nu(1-\lambda\alpha_R/v_F)\) and \(\mathbf{Q}_\lambda=\mathbf{q}/2+\lambda\hat{\mathbf{z}}\times\mathbf{h}/v_F\), and where
\[\Gamma_\lambda=C_\lambda^{(0)}+C_\lambda^{(c)}\cos\theta+C_\lambda^{(s)}\sin\theta\]
are effective vertex functions satisfying Eq. (55-59) in \cite{AttiasKhodas24}. Taking \(\mathbf{q}\perp\mathbf{h}\) for simplicity, we can take \(C_\lambda^{(s)}=0\), and solving these equations to leading order in \(\tau^{-1}\) we find
\begin{widetext}
\begin{align}
    C_{\lambda}^{(0)}&=\lambda\left(1+\frac{1}{4\tau}\sum_{\lambda'}\frac{\nu_{\lambda'}}{\nu}\frac{1}{\sqrt{\omega_n^2+(qv_F/2+\lambda'h)^2}}\right)\\
    C_{\lambda}^{(c)}&=-\frac{i}{4\tau}\sum_{\lambda'}\frac{\lambda'\nu_{\lambda'}}{\nu(qv_F/2+\lambda'h)}\left(1-\frac{|\omega_n|}{\sqrt{\omega_n^2+(qv_F/2+\lambda'h)^2}}\right)\nonumber
\end{align}
\end{widetext}
where we used (assuming \(|a|<|b|\) and \(a>0\)):
\begin{align}
    \int_{0}^{2\pi}\frac{1}{a+i b\cos\theta}\frac{d\theta}{2\pi}&=\frac{1}{\sqrt{a^2+b^2}}\\
    \int_{0}^{2\pi}\frac{\cos\theta}{a+i b\cos\theta}\frac{d\theta}{2\pi}&=i\frac{a-\sqrt{a^2+b^2}}{b\sqrt{a^2+b^2}}\nonumber
\end{align}
We observe that there are two corrections to \(\alpha(q)\) from disorder: one due to the finite electron lifetime in \(\Pi_{\lambda\lambda}\), and another from the correction to the vertex function \(\Gamma\) which implies disorder-induced singlet/triplet mixing.
Undoing the integrals in the previous equation to carry out the Matsubara summation, we then find Eq. \ref{eq:alphaIntraband}.
To evaluate it order by order in \(T^{-1}\), we expand the hyperbolic tangents and evaluate the angular integrals. In particular, we obtain
\begin{widetext}
\begin{align}\label{PiDisI}
    &\iint\frac{\tanh\left(\frac{qv_F/2+\lambda h}{2T}\cos\theta\right)-\tanh\left(\frac{qv_F/2+\lambda' h}{2T}\cos\theta'\right)}{(qv_F/2+\lambda h)\cos\theta-(qv_F/2+\lambda' h)\cos\theta'}d\theta\, d\theta'=\nonumber\\
    &=\sum_{m=1}^\infty\frac{2(2^{2m}-1)}{(2m)!T^{2m-1}}B_{2m}\sum_{k=0}^{m-1}\frac{\Gamma(1/2 + k) \Gamma(-1/2 - k + m)}{\pi k! (m - k-1)!}(qv_F/2+\lambda h)^{2k}(qv_F/2+\lambda' h)^{2m-2-2k}
\end{align}
and
\begin{align}\label{PiDisII}
    &\iint\cos\theta\cos\theta'\frac{\tanh\left(\frac{qv_F/2+\lambda h}{2T}\cos\theta\right)-\tanh\left(\frac{qv_F/2+\lambda' h}{2T}\cos\theta'\right)}{(qv_F/2+\lambda h)\cos\theta-(qv_F/2+\lambda' h)\cos\theta'}d\theta\, d\theta'=\nonumber\\
    &=\sum_{m=1}^\infty\frac{2(2^{2m}-1)}{(2m)!T^{2m-1}}B_{2m}\sum_{k=0}^{m-2}\frac{k+1/2}{k+1}\frac{\Gamma(1/2 + k) \Gamma(-1/2 - k + m)}{\pi k! (m - k-1)!}(qv_F/2+\lambda h)^{2k+1}(qv_F/2+\lambda' h)^{2m-3-2k}
\end{align}
\end{widetext}
where \(\Gamma(x)\) is the gamma function.

Next we evaluate the sums over \(\lambda,\, \lambda'\). We note first that \(\nu_\lambda\nu_{-\lambda}=\nu^2(1-\alpha^2/v_F^2)\approx\nu^2\), and that the linear order in \(h\) term always appears with \(\lambda\) or \(\lambda'\). For \(\lambda=-\lambda'\), therefore, we can set \(h=0\) and evaluate the \(k\) sums in Eqs. (\ref{PiDisI})-(\ref{PiDisII}).
For \(\lambda=\lambda'\) the \(k\) sums can again be done even more trivially. Thus the total contributions from both Eqs. (\ref{PiDisI}) and (\ref{PiDisII}) to linear order in \(h\) and \(\alpha_R/v_F\) is thus found to be
\begin{widetext}
\begin{align}
    \alpha(q)&\approx\dots+\frac{\pi}{2\tau T}\sum_{m=0}^\infty\frac{(2^{2m+2}-1)B_{2m+2}}{(2m+2)!}\left(\nu\left(\frac{qv_F}{2T}\right)+\Delta\nu\left(2-\frac{2\Gamma(m+3/2)}{ \sqrt{\pi}(m+1)!}\right)\left(\frac{2m h}{T}\right)\right)\left(\frac{qv_F}{2T}\right)^{2m-1}
\end{align}

Finally, we use the Taylor series for \(\text{sech}^2\), which can be obtained by differentiating the series for \(\tanh\):
\[\text{sech}^2(x)=\sum_{m=0}^\infty x^{2m}\sum_{k=0}^m \frac{E_{2k}E_{2(m-k)}}{(2k)!(2(m-k))!}\equiv\sum_{m=0}^\infty S_{2m}x^{2m} \]
where
\[S_{2m}=\frac{2^{2m+2}(2^{2m+2}-1)(2m+1)B_{2m+2}}{(2m+2)!}\,,\]
so that in total we have
\begin{align}
    \alpha(q)&=\alpha(q,\tau=\infty)+\frac{\pi}{2\tau T}\sum_{m=0}^\infty\nu\frac{(2^{2m+2}-1)B_{2m+2}}{(2m+2)!}\left(1-\frac{2\Gamma(m+3/2)}{\sqrt{\pi}m!}\right)\left(\frac{qv_F}{2T}\right)^{2m}+\nonumber\\
    &+\frac{\pi}{\tau T}\sum_{m=1}^\infty\Delta\nu\frac{(2^{2m+2}-1)B_{2m+2}}{(2m+2)!}\left(1-\frac{(m+2)\Gamma(m+3/2)}{\sqrt{\pi}(m+1)!}\right)\left(\frac{2m h}{T}\right)\left(\frac{qv_F}{2T}\right)^{2m-1}
\end{align}
The approximate inversion symmetry is broken as long as the coefficients in the sum on the last line are not exactly \(2m\) times the coefficients in the sum on the first line.
\end{widetext}

\bibliography{SDEdisorder}

\end{document}